\documentclass[twocolumn,amsmath,aps]{revtex4}

\usepackage{epsfig}
\usepackage{subfigure}

\newcommand{\be}{\begin{equation}}
\newcommand{\ee}{\end{equation}}
\newcommand{\bea}{\begin{eqnarray}}
\newcommand{\eea}{\end{eqnarray}}

\begin{document}

\title{What the small angle CMB {\it really} tells us about the curvature of
the Universe}
\author{Timothy Clifton}
% \email{tclifton@astro.ox.ac.uk}
\affiliation{Oxford Astrophysics, Physics, DWB, Keble Road, Oxford, OX1 3RH, UK}
\author{Pedro G. Ferreira}
%\email{pgf@astro.ox.ac.uk}
\affiliation{Oxford Astrophysics, Physics, DWB, Keble Road, Oxford, OX1 3RH, UK}
\author{Joe Zuntz}
%\email{jaz@astro.ox.ac.uk}
\affiliation{Oxford Astrophysics, Physics, DWB, Keble Road, Oxford, OX1 3RH, UK}

% ----------------------- ABSTRACT -------------------------

\begin{abstract}
It is well known that observations of the cosmic microwave
background (CMB) are highly sensitive to the spatial curvature of the Universe, $k$.
Here we find that what is in fact being tightly constrained by small angle
fluctuations is spatial curvature near the surface of last scattering,
and that if we allow $k$ to be a function of position, rather
than taking a constant value everywhere, then considerable spatial
curvature is permissible within our own locale.  This
result is of interest for the giant void models that attempt to
explain the supernovae observations without Dark Energy.  We find
voids models with a homogeneous big bang can be compatible with the
observed small angle CMB, but only if they exist in a positively
curved universe.  To be compatible with local measurements of $H_0$, however, we
find that a radially varying bang time is required.
\end{abstract}

\maketitle

% ---------------------- INTRO -------------------------------------------

%\vspace{-10pt}
%\section{Introduction}
%\vspace{-10pt}

One of the great successes of modern cosmology has been the ability of
Cosmic Microwave Background (CMB) anisotropies
to constrain the spatial geometry of the Universe. A succession of ground, sub-orbital
and space-based experiments \cite{cmbexp} have led to increasingly
tight constraints on the curvature of space, $k$, when it is assumed
to be a universal constant.  However, in an
inhomogeneous universe $k$ will not be constant everywhere, but will vary from place
to place. Here we address the question of
what CMB results imply if we allow $k$ to vary with
position. As a result, we place
constraints on models of the Universe in which we live near the centre of a large
under-density, or void. Observables in such models have been
considered previously in \cite{tom}, and have recently been used to explain the supernovae
observations without recourse to Dark Energy  \cite{ellis}-\cite{four}.

Of primary importance for constraining cosmological models are the
$C_\ell$s of the CMB angular power spectrum.  These
quantities are defined by an expansion in Legendre polynomials,
$P_\ell(x)$, of the form  $\langle \delta T({\bf n})\delta T({\bf n'})\rangle=\frac{1}{4\pi}
\sum_\ell (2\ell+1)C_\ell P_\ell({\bf n}\cdot{\bf n'})$, where $\delta T({\bf n})$ is
the CMB temperature anisotropy in the direction ${\bf n}$, and angled
brackets indicate an ensemble average. Here we will
focus on the properties of the $C_\ell$s on
small angular scales. They are then a result of two processes:  The imprint of cosmological perturbations onto the last
scattering surface, and the projection of that surface onto our sky.

The first of these processes occurs early enough in the Universe's
history that it is relatively insensitive to the effects of any
spatial curvature.  It can then be accurately described by linear
perturbation theory about a flat background. The familiar set of peaks
and troughs in the $C_\ell$s are then determined by cosmological
parameters such as the expansion rate up to last scattering, and
the relative densities of the different constituents of the Universe \cite{JKS}.

The second process involves
relating length scales at last scattering to angles on the sky today,
and is highly sensitive to the geometry of the intervening
space-time. Indeed, it is well
known that non-zero $k$ results in a shift of the
acoustic power spectrum of small scale fluctuations in the
CMB \cite{GSS}, and that it is this effect that is responsible
for the stringent constraints on spatial curvature that usually imply
$k \sim 0$.  Such constraints, however, assume that $k$ is a constant, throughout
the Universe.  Here we relax this condition, and allow $k$ to vary with
position, by considering the spherically symmetric
Lema\^{i}tre-Tolman-Bondi (LTB) space-time. We find that $k({\bf x})$ is only well constrained in the
vicinity of the surface of last scattering, and that even large local fluctuations
in $k$ will only produce moderate contributions to the shift.

On small angular scales, the relative temperature of the CMB seen on an observer's sky in different directions, $\hat{{\bf n}}$,
is given by ${\delta T}(\hat{{\bf n}}) = \Delta(\hat{{\bf n}} D_{LS})$,
where $D_{LS}$ is some measure of the distance to the last scattering
surface and $\Delta$ is a solution of the Einstein-Boltzmann equations at the time of
last scattering.  In conformally static space-times, such as those
  with $k=$constant,
$D_{LS}$ can unambiguously be taken to be the conformal
distance to last scattering $r=\sinh \left( \sqrt{-k} \int d \eta \right)/\sqrt{-k}$, where $d \eta \equiv
dt/a$ is conformal time. For more general space-times, however, we
will need to be more careful. 

Assuming that the radius of curvature is much greater than the scale
of any perturbations, we have that the variance in temperature
fluctuations is
$\left\langle \delta T(\hat{{\bf n}}) \delta T(\hat{{\bf n}}^{\prime})
\right\rangle
=  \int d^3k \mathcal{P}_{\Delta}({\bf k}) e^{i
  {\bf k} \cdot (\hat{ {\bf n} }-\hat{ {\bf n} }^{\prime}) D_{LS}}$,
where we have defined the power spectrum to be $\mathcal{P}_{\Delta}(|{\bf k}|)\delta^3({\bf k}-{\bf k}')\equiv\langle \Delta^*({\bf
  k}) \Delta({\bf k}^{\prime}) \rangle $.     
What we are ultimately interested in is the
angle between vectors ${\bf \hat{n}}$ and ${\bf
  \hat{n}^{\prime}}$ at the observer, $d \theta = \cos^{-1} (\hat{{\bf n}} \cdot  \hat{{\bf n} }^{\prime} )
\simeq \vert   \hat{ {\bf n} }-\hat{ {\bf n} }^{\prime}  \vert$,
hence, if $dp= \vert {\bf \hat{n}}- {\bf \hat{n}^{\prime}} \vert D_{LS}$ is the
distance between two points at last scattering, then
%\begin{equation}
$D_{LS} = dp/d\theta \equiv d_{A,LS}$,
%\end{equation}
where $d_{A,LS}$ is the angular diameter distance to last
scattering. This is a generic result valid for any curvature, constant or not.
We will now approximate the $C_\ell$s as a Fourier decomposition of
$\left\langle \delta T(\hat{{\bf n}}) \delta T(\hat{{\bf n}}^{\prime})
\right\rangle$ over the sky. Defining
the two dimensional wave number, ${\bf q}$, such that
$q\equiv|{\bf q}|=\ell$, we then have $C_\ell\simeq C_q$, where
%\begin{eqnarray}
\be
C_q = \int d \Omega^2
\left\langle \delta T (\hat{{\bf n}})  \delta T (\hat{{\bf n}}^{\prime})
\right\rangle e^{-i {\bf q}\cdot {\bf \theta}} \nonumber 
= \frac{1}{d^2_{A,LS}} \mathcal{P} _{\Delta}\left( \frac{\vert {\bf q}
\vert }{d_{A,LS}}
\right).
\label{cl}
\ee
%\end{eqnarray}
On small enough scales, of a few degrees and below, we expect this expression to be
good enough for accurate parameter estimation \cite{baos}.
%\footnote{This is
%  not true of Baryon Acoustic Oscillations (BAOs) when
%  the Universe is inhomogeneous. The curvature in the background will 
%affect the position and amplitude of the oscillations in the galaxy power 
%spectrum in a far more complicated way. Hence we do not use BAOs.}.

\begin{figure}[tbp]
\begin{flushright}
\vspace{-5pt}
\epsfig{figure=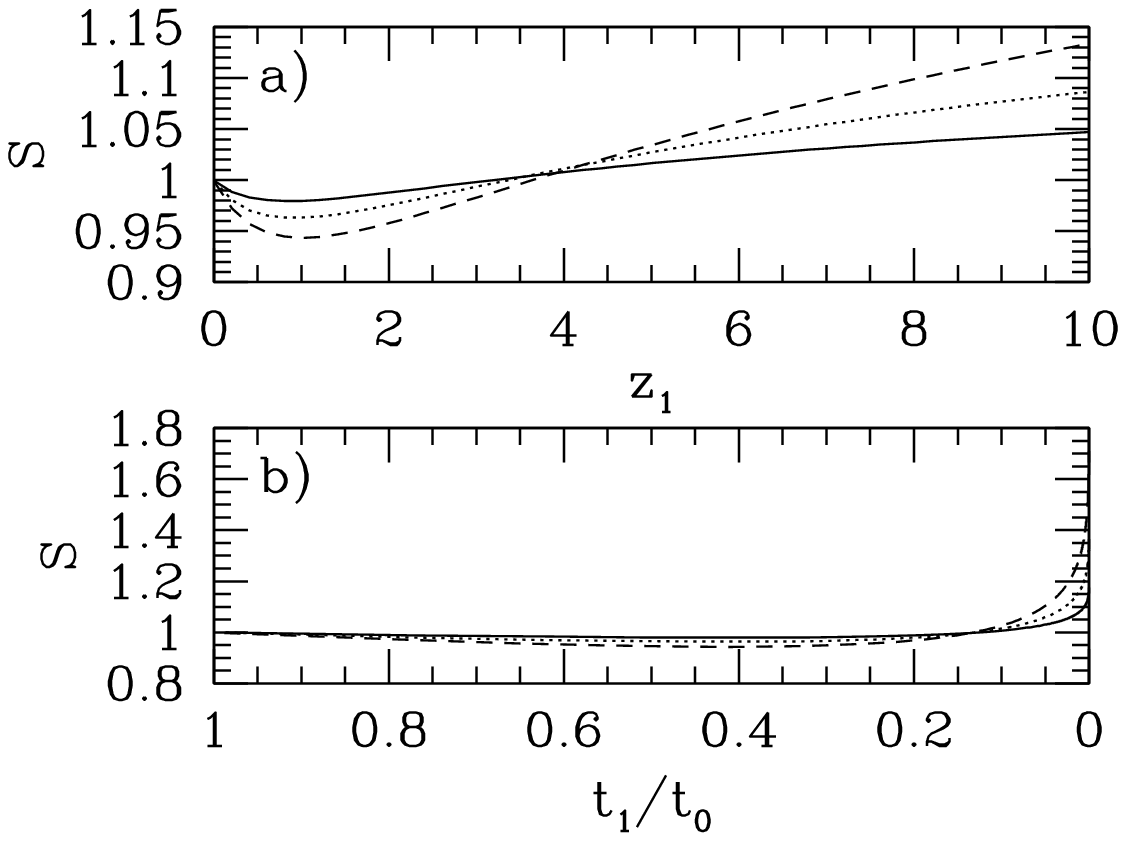,width=8.4cm}
%\subfigure{\epsfig{figure=Figure1b.eps,width=8.4cm}}
\end{flushright}
\vspace{-20pt}
\caption{Upper panel: The shift parameter, $S$, as a function of $z_1$ for
  $\Omega_K=0.3$ (solid), $0.5$ (dotted) and $0.7$ (dashed). Lower
  panel: $S$ as
  a function of cosmic time, $t_1=t(z_1)$, for the same models ($t_0$
  is time today).}
\label{shift}
\vspace{-20pt} 
\end{figure}

Now consider two different space-times.  Although it can be arranged
that observers in each of these will witness identical last scattering
surfaces at identical redshifts, the geometries
between those observers and that surface will be different in each.
Let us write $d_{A,LS}$ for the angular diameter distance in the first space-time,
and $\hat{d}_{A,LS}$ for the angular diameter distance in the second. 
We can then relate the angular power spectrum
in the first space-time, $C_{\ell}$, to that in the
second space-time, $\hat{C}_{\ell}$, via
$ C_{\ell} = S^2
\hat{C}_{\ell/ S}$,
where $S \equiv \hat{d}_{A,LS}/d_{A,LS}$ is known as the shift parameter. 
This situation (of identical last scattering surfaces but different geometries) is often envisaged when considering the effect of a
non-zero, and constant, $k$.  In that case both space-times are
conformally static, and so it suffices to use the
conformal (or optical) metric.  The effect of $k \neq 0$ is then to
alter the conformal distance to the last scattering surface, and the
$C_{\ell}$s of each observer can be related by a shift parameter that
is the ratio of these conformal distances \cite{KSS}.  

Now consider a toy model with a region of curved FRW extending out to some
redshift, $z_1$, in a universe that is otherwise flat.  We have in
Friedmann-Robertson-Walker (FRW) cosmology that the angular diameter
distance is given by $d_A=a r$, where $r$ is
conformal distance (defined above) and $a$ is the scale factor of the
universe. A dust-filled FRW universe can be shown to have 
$d_A$, as a function of the redshift $z\equiv a_0/a-1$, given by 
%\bea
%\label{dA}
%d_A &= \frac{1}{(1+z) H_0 \sqrt{\Omega_k}} \sinh \Bigg\{ 2 \sinh^{-1}
%  \sqrt{\frac{\Omega_k}{(1-\Omega_k)}} \qquad \qquad \qquad \\ &\qquad \qquad \qquad \qquad -  2 \sinh^{-1} \sqrt{
%    \frac{1}{(1+z) }\frac{\Omega_k}{(1-\Omega_k)}} \Bigg\} ,
%\nonumber
%\eea
\be
\label{dA}
d_A = \frac{\sinh (2 \hat{r})}{(1+z) H_0 \sqrt{\Omega_k}},
\ee
where $\hat{r}\equiv \sinh^{-1}
  \sqrt{\frac{\Omega_k}{(1-\Omega_k)}}  -  \sinh^{-1} \sqrt{
    \frac{1}{(1+z) }\frac{\Omega_k}{(1-\Omega_k)}}$,
$H$ is the Hubble rate, subscript $0$ denotes a quantity measured by the observer at $z=0$, and
$\Omega_k \equiv -k/a_0^2 H_0^2$.   The shift in CMB
peaks from a globally flat universe is now given by the ratio
$d_A^{\text{curved}}/d_A^{\text{flat}}$ at $z_1$, when $H$ has been
matched at last scattering (and so is also matched at $z_1$). In a flat universe
we have $H_{0,\text{flat}}^2=H^2_{z}/(1+z)^3$, where $H_{z}$ is the value of
$H$ at redshift $z$, and in a spatially curved
universe we have $H_{0,\text{curved}}^2 ( 1-\Omega_k z/(1+z))= H^2_{z}/(1+z)^3$.
This then gives the shift parameter as
%To calculate the shift parameter we now need the ratio of present day Hubble
%rates in each space-time, given that each observer is witnessing a
%surface with the same Hubble rate. $H_0$ in a flat universe is given by
%$H_{0,\text{flat}}^2=H^2_z/(1+z)^3$, where $H_z$ is the value of
%$H$ at redshift $z$, and in a spatially curved
%universe we have $H_{0,\text{open}}^2 ( 1-\Omega_k z/(1+z))= H^2_z/(1+z)^3$.
%The shift parameter between flat and curved FRW universes is then
%\bea
%\label{SFRW}
%S(z) &=
%    \frac{\sqrt{(1+z)-z \Omega_k}}{2 \sqrt{\Omega_k} (\sqrt{1+z}-1)} \sinh \Bigg\{ 2 \sinh^{-1}
%  \sqrt{\frac{\Omega_k}{(1-\Omega_k)}} \qquad \qquad \qquad \\ &\qquad \qquad \qquad \qquad -  2 \sinh^{-1} \sqrt{
%    \frac{1}{(1+z) }\frac{\Omega_k}{(1-\Omega_k)}} \Bigg\}.
%\nonumber
%\eea
\be
\label{SFRW}
S(z_1) =
    \frac{\sqrt{(1+z_1)-z_1 \Omega_k}}{2 \sqrt{\Omega_k} (\sqrt{1+z_1}-1)}
    \sinh (2 \hat{r}_1).
\ee

In Fig. \ref{shift} we plot this shift as a
function of $z_1$ and the corresponding cosmic time for
three choices of curvature, $\Omega_k=0.3$, $0.5$ and $0.7$.  At large $z_1$ we
recover the familiar result that $k<0$ leads to $S>1$, so that the
acoustic peaks of the CMB are shifted to smaller angular scales.
However, if we consider curved regions out to lower redshift, then this result is no longer true:
At $z_1 \lesssim 4$ negative curvature causes $S<1$.  This is ultimately
due to the presence of $H_0$ in Eq. (\ref{dA}).  Measuring $d_A$ in units of
$h^{-1}$, $S$ would increase monotonically with $z$.  The requirement that 
$H_{LS}$ is the same in both space-times, however, leads to
different values of $H_0$ in each.  At low redshifts the ratio of
these Hubble rates is great enough to cancel what would otherwise be a
positive $S-1$.

In the lower panel of Fig. \ref{shift} we plot $S$ all the way out to
last scattering, now as a function of cosmic time, $t_1=t(z_1)$, in the
fiducial flat model.  Here it can be seen that most
of the shift parameter is due to geometrical effects shortly
after the surface of last scattering, at $z_*\simeq1100$, with any
effects due to our local geometry contributing significantly less.  In fact, for $\Omega_k \sim 0.7$ it can be
seen that there is only a $\sim 5\%$ shift caused by all of the geometry out
until the Universe was $\sim 5\%$ of its current age.  The rest of the
$\sim 70\%$ shift at last scattering is then primarily due to the geometry
experienced by the CMB photons in the first $\sim 5\%$ of the Universe's history.

%\vspace{-10pt}
%\section{Probing local curvature in Lemaitre-Tolman-Bondi Universes.}
%\vspace{-10pt}

Now let us consider models in which $k$ is a smoothly varying function
of position, as emerges in a universe with 
large density fluctuations \cite{sylos}.
To achieve this consider the LTB model, whose line-element is given by
 \cite{LTB1}
\be
\label{LTB}
ds^2 = -dt^2 + \frac{a_2^2(t,r) dr^2}{1-k(r) r^2} + a_1^2(t,r) r^2 d \Omega^2,
\ee
where $a_2 = (r a_1)^\prime$, and primes denote
partial derivatives with respect to $r$.  The FRW scale factor, $a$,
has now been replaced by two new scale factors, $a_1$ and $a_2$, describing expansion in
the directions tangential and normal to surfaces of spherical
symmetry.  These new scale factors are functions of cosmic time, $t$, and
distance, $r$, from the centre of symmetry, and obey a generalization
of the usual Friedmann equation such that
\be
\label{Fried}
\left(\frac{\dot a_1}{a_1}\right)^2 = \frac{8\pi G}{3} \frac{m(r)}{a_1^3} -\frac{k(r)}{a_1^2},
\ee
where over-dots are partial derivatives with respect to $t$.
The energy density is given here by $\rho = (mr^3)^\prime/ a_2 a_1^2 r^2$, and
redshifts by $1+z = \text{exp} \{ \int (\dot{a}_2/a_2) dt \}$,
where the integral is along a past directed radial null geodesic.
\begin{figure}[tbp]
%\begin{center}
%\vspace{-25pt}
\epsfig{figure=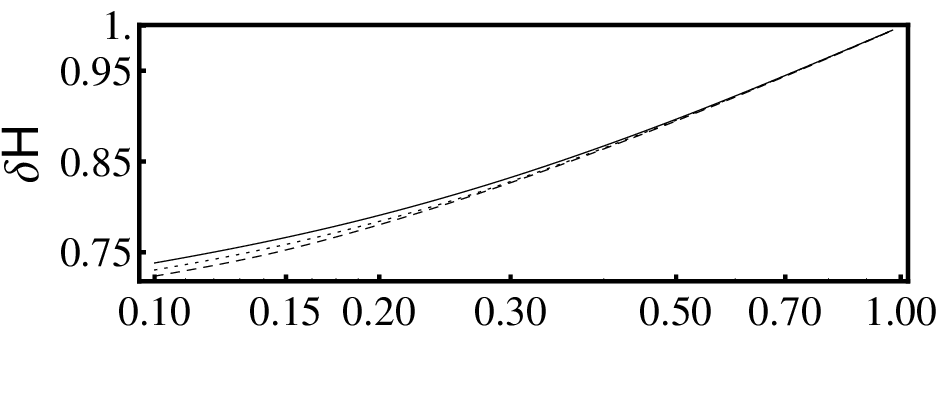,width=8cm}
\epsfig{figure=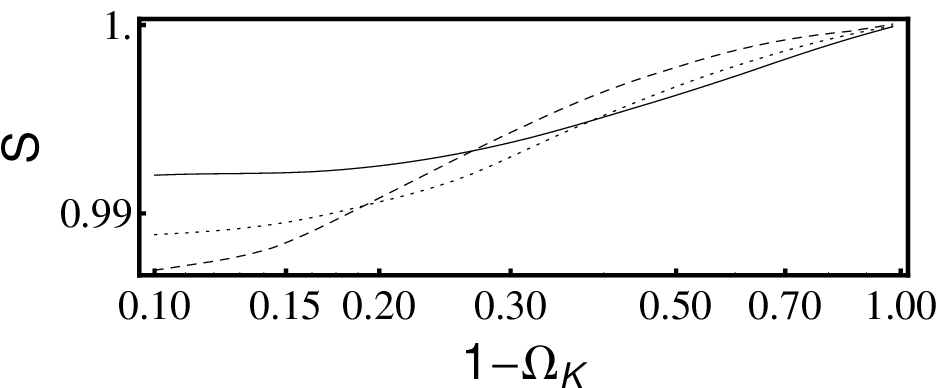,width=8cm}
%\subfigure{\epsfig{figure=Figure1b.eps,width=8.4cm}}
%\end{center}
\vspace{-5pt}
\caption{Upper panel: $\delta H$ (see text for definition) as a function of
central void curvature, $\Omega_K$, for asymptotically flat voids with FWHM at redshift
of 0.4 (solid), 0.5 (dotted) and 0.6 (dashed). Lower panel: The shift
parameter, $S$, as a function of central void curvature, $\Omega_K$,
for the same three voids. }
\label{voidshift}
\vspace{-15pt}
\end{figure}

The LTB space-time is fully determined by a choice of the three free
functions $k(r)$, $m(r)$ and $t_0(r)$.  The first two of these are
specified above, and the third is the `bang time', which in these
models need not be the same at all points in space.  Without loss of
generality, we can then make a coordinate choice such that $m=$constant.
We will also initially consider the situation of a simultaneous big bang, with $t_0=$constant.  These
models have been much studied recently, as a space-time with local
negative curvature allows for the possibility of explaining the
supernova data without Dark Energy.  A fit to the
data is often found to be a void with $\Omega_k  \sim 0.7$, and a
width of $z \sim 0.5$.  This is a significant amount of spatial
curvature, extending out to large distances, and one may 
naively suspect that the sensitivity of the small angle CMB to
spatial curvature may be sufficient to impose strong constraints on
these models \cite{CMBvoid1}-\cite{zibinCMB}. 

To investigate if this is indeed so, let us consider a negative local curvature fluctuation
in an otherwise flat universe.  An observer at the
centre of such a void will see a last scattering surface at $z_*$,
and can straightforwardly calculate $H$ at this surface in terms of
their locally measured value.  We also require a fiducial observer in
an FRW universe who will witness an identical last scattering surface,
with the same $H_{LS}$.  To ensure that these observers use comparable
measures of distance we will enforce the conditions that they have the
same local geometry \cite{bonnor}.
% \footnote{For non-singular voids it can be shown
%  that the geometry at the centre is locally Friedman. See Bonnor, W. {\it MNRAS} {\bf
%  167}, 55 (1974).}.  
This choice ensures that distances to nearby
co-moving objects are the same when measured in units of $h^{-1}$MPc. We also require that they both see
last scattering surfaces at the same $z_*$ so that effects due to
the redshifting of solid angle, for example, are automatically included. 

%\begin{figure}[htbp]
%\begin{center}
%\vspace{-4pt}
%\epsfig{figure=Fig3.eps,width=8.7cm}
%\subfigure{\epsfig{figure=Figure1b.eps,width=8.4cm}}
%\end{center}
%\vspace{-15pt}
%\caption{The $68\%$ and $95\%$ confidence regions from $l>100$ on the central curvature, $\Omega_K$, and FWHM
%  of $k$ in redshift, $z_E$, of an asymptotically flat void with simultaneous
%  big bang.  Shell crossing occurs in the excluded hatched region.}
%\label{voidomegaz}
%\vspace{-17pt}
%\end{figure}

The shift between the open FRW universe and the void model is then given by
$S_1=d_{A,LS}^{\text{LTB}}/d_{A,LS}^{\text{open}}$, where the angular
diameter distance in LTB is given by $d_{A,LS}^{\text{LTB}}=a_{1,LS}
r_{LS}$, and in the FRW universe by Eq. (\ref{dA}).  
 In the case of the void model, the values
of $a_1$, $r$ and $H$ at last scattering are found by
integrating a radial null geodesic out to $z_*$, using the
solutions to Eq. (\ref{Fried}).  $H_0$ in the open FRW universe
is then found by taking the same Hubble rate at last scattering as in the LTB
model, and propagating it forward until today in the FRW geometry.
Of course, we know the shift parameter between open 
and flat FRW universes,
$S_2=d_{A,LS}^{\text{open}}/d_{A,LS}^{\text{EdS}}$ from (\ref{SFRW}), and so we can calculate the
acoustic spectrum witnessed by the observer in the void in terms of a
shift, $S=S_1 S_2=d_{A,LS}^{\text{LTB}}/d_{A,LS}^{\text{EdS}}$, from a spatially flat FRW model, and a
change in Hubble rate, $\delta H \equiv H_0^{EdS}/H_0^{LTB}$.

The shift, $S$, and change in Hubble rate, $\delta H$, for an asymptotically flat void formed from a
negative Gaussian perturbation in $k(r)$, are shown in
Fig. \ref{voidshift}. We find that a good fit to the WMAP data requires $S \sim 0.9$ and $\delta H \sim
0.5$, and so a void model will need to be capable of achieving similar
values if it is to be considered viable.  It can immediately be seen
that for moderately deep voids, with $\Omega_k \lesssim 0.9$ at the centre, both $S$ and
$\delta H$ deviate insufficiently from $1$ \cite{afterbonnor}.
% \footnote{This is due to the value
% of $S_2>1$, from Fig. \ref{shift}, being largely cancelled by the
% value of $S_1<1$, leaving $S=S_1 S_2 \sim 1$.}.
It can also be seen that $S$ is not particularly sensitive to the width of the void.
In light of what we considered above, these results
can be easily understood:  Most of the contribution to the
shift does not come from the local geometry, but from early times when
the CMB photons were well outside the void. The discussion above also
explains why the presence of the void shifts the acoustic peaks to
larger scales, rather than smaller.

One may also wish to consider more extreme voids in which
we allow $\Omega_k > 0.9$ at the centre.  
%In this case both $S$ and $\delta H$
%can deviate considerably from $1$.  We have used a Monte Carlo Markov Chain
%algorithm running a modified version of the publicly available 
%CosmoMC \cite{CAMB} to find constraints on $\Omega_K$ and
%the extent of the void, $z_E$ from the WMAP 5 year data, with
%$l>100$. 
In this case, however, shell crossing singularities can occur
\cite{shell}, and redshift as a function of local energy density can become
multi-valued \cite{zibin2}.  The former of these should be considered as a
break-down of the model, while the latter shows that the effect of the
inhomogeneity on null geodesics in these cases can be highly non-trivial.
% \footnote{Shell crossing
%  occurs when $a_2=0$, and indicates a breakdown of the model due to
%  the neglect of pressure in the matter content.} 
%we find the best fit void has 
%$\Omega_k=0.986^{\text{\tiny $+0.002$}}_{\text{\tiny $-0.001$}}$ at
%its centre, when we consider voids with a FWHM in $k(r)$ at less than
%$z=1.2$.  The likelihood contours in this parameter space are shown in Fig. \ref{voidomegaz}.
%The best fit model in this region of parameter space is effectively identical to
%$\Lambda$CDM, with $\Delta \chi^2 = 0.4$.  This success at fitting the
%detailed shape of the CMB adds some perspective to claims that the
%CMB tightly constrains the geometry of the Universe and provides
%evidence for Dark Energy (when accompanied by an independent measure
%of $H_0$).  In performing this fit we have used the Hubble key
%project constraint $h=0.72\pm 0.08$
%\cite{HST}.  The fit to the large $l$ CMB then gives us the combined
%results for local Hubble rate  $h=0.70\pm 0.01 $, and baryon fraction
%$\Omega_b h^2= 0.07^{\text{\tiny $+0.01$}}_{\text{\tiny $-0.03$}}$.
However, even if one is prepared to consider such extreme voids, and even if
they can be made compatible with the small angle CMB, such voids are
still highly unlikely to be able to fit the supernova data without having their
shape at low $z$ being highly fine-tuned.  We will not consider them
further here.

\begin{figure}[tbp]
\begin{center}
\vspace{-10pt}
\epsfig{figure=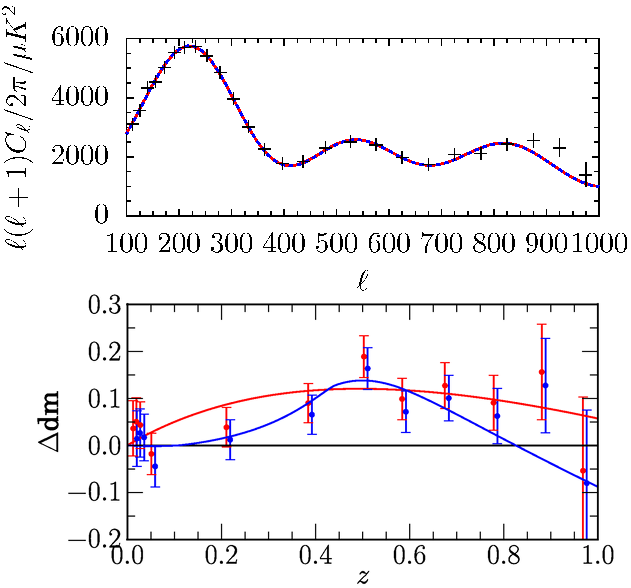,width=8.4cm}
%\subfigure{\epsfig{figure=Figure1b.eps,width=8.4cm}}
\end{center}
\vspace{-15pt}
\caption{Top panel: the $C_{\ell}$s  for the void in a
  non-flat background (blue) and for $\Lambda$CDM with $\Omega_{\Lambda}=0.7$ (red) are essentially indistinguishable.  
  Bottom panel: the distance modulus for the same two models. Data
  points are from the WMAP 5 year data \cite{cmbexp} (top panel) and
  SNLS first year data \cite{SN} (bottom panel); in the latter case, the position of
  the data points move as they are fitted to the two models (see \cite{clifton}).}
\label{sne}
\vspace{-15pt}
\end{figure}

Of course, one will still be interested in more general void models.  In
particular, it is possible to conceive of a void in a spatially curved
FRW universe, instead of a flat one.  In this case one is subject to
the familiar sensitivity of the CMB to spatial curvature, and we have
verified that $S$ can effectively be set to any value with a suitable
choice of asymptotic curvature \cite{zibinPC}.  In particular, a shift
parameter of $S \simeq 0.9$ can be achieved with $\Omega_k \simeq
-0.26$ asymptotically.  The value of $\delta H$, however, is
not so sensitive to $k$ in the background space-time.  To achieve
$\delta H \sim 0.5$ one must therefore be prepared to abandon the notion of the
big bang happening at the same time at all points in space. A larger
contrast between local and asymptotic Hubble rates can then be
straightforwardly achieved.  To this end, we find that a Gaussian void
embedded in a spatially curved universe with $\Omega_k=0.10$, that has a FWHM in $k$ at
$z=0.34$, in $t_0$ at $z=0.80$, and with an age of the
universe in the centre of the void that is $13\%$ more than that of the asymptotic
regions, we can fit the $C_{\ell}$s just as well as $\Lambda$CDM with
$\Omega_m=0.15$ at the centre of the void.  The CMB
acoustic spectrum and distance modulus plot for this void are shown in
Fig. \ref{sne}, together with the $\Lambda$CDM best fits. Changing the
detailed shape of the under-density will change the numbers involved
above, and, in particular, if one can find other voids that 
allow $\delta H \sim 0.5$ then these models will very likely provide a good
fit to the data too (with the appropriate choice of background curvature,
to give the correct shift).

In conclusion, we find that the observed acoustic spectrum of
small angle CMB fluctuations is primarily only sensitive to curvature at
high redshifts.  Local curvature has much smaller, and even opposite,
effects.  By considering LTB models, in which $k=k(r)$, we demonstrate that large local fluctuations in
spatial curvature produce only moderate shifts in the CMB acoustic
spectrum.  As a result, the local void models that
seek to explain cosmological observations without Dark
Energy are not automatically ruled out.  Fitting to the WMAP 5 year data shows, however, that
the simplest voids (with simultaneous big bang) are required to be
have non-zero asymptotic spatial curvature.  By embedding the
void in a suitably curved background it is then possible to shift the
acoustic spectrum by any amount.  Even in this case,
however, the locally observed Hubble rate in the void model is
anomalously low.  Alternatively, we can give up on the idea of a
simultaneous big bang.   In this case it is found that the local
Hubble rate (as well as the shift parameter) is
sensitive to the bang time function, and by altering the age of the
Universe in different spatial locations we can increase $H_0$.  We
therefore find void models that can fit the WMAP 5 year data just as
well as $\Lambda$CDM, as well as local measurements of $H_0$, and
supernova observations.  However,
if we really do live in a large, local under-density in the Universe,
it will have to be considerably more complex than previously thought in
order to be observationally viable.

%\vspace{-10pt}
%\section{Discussion: is there a viable LTB model?}
%\vspace{-10pt}

% ------------------------ ACKNOWLEDGEMENTS ----------------------------------
\vspace{-2pt}
\section*{Acknowledgements}
\vspace{-10pt}

We are grateful to P. Bull, C. Clarkson, R. Durrer , J. Silk and J. Zibin for
helpful comments, and to the BIPAC for support.  TC acknowledge the
support of Jesus College, and JZ that of the STFC.

% ---------------------- BIBLIOGRAPHY -----------------------------------------
%\vspace{-20pt}


\begin{thebibliography}{100}
\bibitem{baos} This is not true of Baryon Acoustic Oscillations (BAOs)
  when the Universe is inhomogeneous. The curvature in the background
  will affect the position and amplitude of the oscillations in the
  galaxy power spectrum in a far more complicated way. Hence we do not use BAOs.
\bibitem{bonnor} For non-singular voids it can be shown
  that the geometry at the centre is locally Friedman. See Bonnor, W. {\it MNRAS} {\bf
  167}, 55 (1974).
\bibitem{afterbonnor} This is due to the value of $S_2>1$, from
  Fig. \ref{shift}, being largely cancelled by the  value of $S_1<1$,
  leaving $S=S_1 S_2 \sim 1$.
\bibitem{zibinPC} This may explain how the ``unconstrained'' models of
\cite{zibinCMB} were able to produce a suitable fit to the data, albeit with
low $H_0$, as these models have a region of $k>0$ outside
the central under-density.
\end{thebibliography}
\end{document}